\documentstyle[12pt]{article}
\makeatletter
\def\maketitle{\par
 \begingroup
 \def\thefootnote{\fnsymbol{footnote}}
 \def\@makefnmark{\mbox{$^\@thefnmark$}}
 \@maketitle
 \@thanks
 \endgroup
 \setcounter{footnote}{0}
 \let\maketitle\relax
 \let\@maketitle\relax
 \gdef\@thanks{}\gdef\@author{}\gdef\@title{}\let\thanks\relax}
\def\@maketitle{\vspace*{0.9cm}
{\hsize\textwidth
 \linewidth\hsize \centering
 {\normalsize \bf \@title \par} \vskip 1.0cm  {\normalsize  \@author \par}}}

\def\thefootnote{\mbox{\noindent$\fnsymbol{footnote}$}}
    \long\def\@makefntext#1{\noindent$^{\@thefnmark}$#1}

\def\section{\@startsection{section}{1}{\z@}{1.5ex plus 0.5ex minus
   1.2ex}{1.3ex plus .1ex}{\normalsize\bf}}
\def\subsection{\@startsection{subsection}{2}{\z@}{1.5ex plus 0.5ex minus
    1.2ex}{1.3ex plus .1ex}{\normalsize\em}}

\def\@sect#1#2#3#4#5#6[#7]#8{\ifnum #2>\c@secnumdepth
     \def\@svsec{}\else
     \refstepcounter{#1}\edef\@svsec{\ifnum #2=1 \@sectname\fi
        \csname the#1\endcsname.\hskip 1em }\fi
     \@tempskipa #5\relax
      \ifdim \@tempskipa>\z@
        \begingroup #6\relax
          \@hangfrom{\hskip #3\relax\@svsec}{\interlinepenalty \@M #8\par}
        \endgroup
       \csname #1mark\endcsname{#7}\addcontentsline
         {toc}{#1}{\ifnum #2>\c@secnumdepth \else
                      \protect\numberline{\csname the#1\endcsname}\fi
                    #7}\else
        \def\@svsechd{#6\hskip #3\@svsec #8\csname #1mark\endcsname
                      {#7}\addcontentsline
                           {toc}{#1}{\ifnum #2>\c@secnumdepth \else
                             \protect\numberline{\csname the#1\endcsname}\fi
                       #7}}\fi
     \@xsect{#5}}

\def\@sectname{}

\def\thebibliography#1{\section*{{{\normalsize
\bf References }
\rule{0pt}{0pt}}\@mkboth
  {REFERENCES}{REFERENCES}}\list
  {{\arabic{enumi}.}}{\settowidth\labelwidth{{#1}}%
    \leftmargin\labelwidth  \frenchspacing
    \advance\leftmargin\labelsep
    \itemsep=-0.2cm
    \usecounter{enumi}}
    \def\newblock{\hskip .11em plus .33em minus -.07em}
    \sloppy
    \sfcode`\.=1000\relax}

\def\@cite#1#2{\unskip\nobreak\relax
    \def\@tempa{$\m@th^{\hbox{\the\scriptfont0 #1}}$}%
    \futurelet\@tempc\@citexx}
\def\@citexx{\ifx.\@tempc\let\@tempd=\@citepunct\else
    \ifx,\@tempc\let\@tempd=\@citepunct\else
    \let\@tempd=\@tempa\fi\fi\@tempd}
\def\@citepunct{\@tempc\edef\@sf{\spacefactor=\the\spacefactor\relax}\@tempa
    \@sf\@gobble}

\def\citenum#1{{\def\@cite##1##2{##1}\cite{#1}}}
\def\citea#1{\@cite{#1}{}}

\newcount\@tempcntc
\def\@citex[#1]#2{\if@filesw\immediate\write\@auxout{\string\citation{#2}}\fi
  \@tempcnta\z@\@tempcntb\m@ne\def\@citea{}\@cite{\@for\@citeb:=#2\do
    {\@ifundefined
       {b@\@citeb}{\@citeo\@tempcntb\m@ne\@citea\def\@citea{,}{\bf ?}\@warning
       {Citation `\@citeb' on page \thepage \space undefined}}%
    {\setbox\z@\hbox{\global\@tempcntc0\csname b@\@citeb\endcsname\relax}%
     \ifnum\@tempcntc=\z@ \@citeo\@tempcntb\m@ne
       \@citea\def\@citea{,}\hbox{\csname b@\@citeb\endcsname}%
     \else
      \advance\@tempcntb\@ne
      \ifnum\@tempcntb=\@tempcntc
      \else\advance\@tempcntb\m@ne\@citeo
      \@tempcnta\@tempcntc\@tempcntb\@tempcntc\fi\fi}}\@citeo}{#1}}
\def\@citeo{\ifnum\@tempcnta>\@tempcntb\else\@citea\def\@citea{,}%
  \ifnum\@tempcnta=\@tempcntb\the\@tempcnta\else
   {\advance\@tempcnta\@ne\ifnum\@tempcnta=\@tempcntb \else \def\@citea{--}\fi
    \advance\@tempcnta\m@ne\the\@tempcnta\@citea\the\@tempcntb}\fi\fi}

\def\abstract{\if@twocolumn
\section*{Abstract}         
\else \small
\begin{center}
{ABSTRACT\vspace{-.5em}\vspace{0pt}}
\end{center}
\quotation
\fi}
\def\endabstract{\if@twocolumn\else\endquotation\fi}

\def\fnum@figure{Fig. \thefigure}

\long\def\@makecaption#1#2{
   \vskip 10pt
   \setbox\@tempboxa\hbox{\small #1. #2}
   \ifdim \wd\@tempboxa >\hsize    
      \small #1. #2\par            
   \else                           
      \hbox to\hsize{\hfil\box\@tempboxa\hfil}
   \fi}
\makeatother

\begin{document}
\rightline{PSU/TH/132}
  \title{PHYSICS OF POLARIZED pp COLLISIONS}

\author{John C. Collins\thanks{E-mail: collins@phys.psu.edu}
     \thanks{To appear in
      Proceedings of PANIC XIII --- Particles and Nuclei International
      Conference at Perugia, Italy 28 June -- 2 July 1993 (World
      Scientific, Singapore, 1993).}
  \\
     {\em
        Penn State University, 104 Davey Lab, University Park,
     \\
        PA 16802, U.S.A.
     }
  }
  \maketitle
  \setlength{\baselineskip}{2.6ex}

\begin{abstract}

    I will summarize the physics that can be investigated with
    polarized pp collisions. It is technically feasible to use
    the RHIC collider for accelerating highly polarized protons
    to a center-of-mass energy of about 400 GeV, with high
    luminosity. Such collisions can be used to probe the
    spin-dependence of hard collisions and of partons in a
    hadron, including the gluons. There are interesting twist 3
    phenomena that are likely to be significant in view of the
    large transverse spin asymmetries at lower energies; this
    will have important implications for the spin structure of
    the proton wave function. Recent theoretical developments
    include the possibility of probing the spin of transversely
    polarized quarks via asymmetries in the jets they make.

\end{abstract}

\section{Introduction}

It has been realized that it is possible to accelerate polarized
protons in the Relativistic Heavy Ion Collider (RHIC) at
Brookhaven, at a modest incremental cost. This opens up to
reality a whole area of QCD physics. Therefore a collaboration
has been formed---the RHIC Spin Collaboration (RSC) to promote
this kind of physics \cite{RSC}.

The polarized RHIC would have proton beams of up to 250 GeV of
energy, with $70 \%$ polarization and a luminosity of around
$10^{32}{\rm cm}^{-2}{\rm sec}^{-1}$, and we would propose running in
polarized mode for 1 month a year.  This would clearly allow hard
scattering physics to be done with polarization under conditions
comparable to those at proton-antiproton colliders.

The technical development that has allowed this is known as the
Siberian snake \cite{snake}.
This is a spin rotator that does not affect the
beam optics and can be used to cancel most of the depolarizing
resonances. Previously acceleration of polarized protons to high
energy was a tour-de-force, even at the modest energy of 20 GeV
at the Brookhaven AGS. Snakes would make this easy. Tests of the
snake concept have been made at low energy at the Indiana
cyclotron \cite{snaketest}, while tests at higher energy
are underway at the AGS this year.  (The AGS will
serve as the injector for RHIC).

Compared to previous experiments, RHIC would have high energy,
luminosity and undiluted high polarization, with easy spin
reversal.  Hence it will allow us to do high quality
short-distance QCD physics with polarization effects.  Among
other things, this will give a good probe of the spin structure
of the proton wave function, which in turn will provide a much
needed window onto chiral symmetry and its breaking.


\section{Polarized Hard Scattering in QCD}

QCD predicts that suitable cross sections are a convolution of
parton densities (and fragmentation functions, if needed) and of
short-distance parton cross sections.  Such cross sections
include jet production, the Drell-Yan process, heavy flavor
production, direct photon production at large transverse momentum.

In collisions of longitudinally polarized beams, there is an
asymmetry of the cross section when one of the beam helicities is
reversed.  Schematically the asymmetry in the cross section is of
the form
\begin{equation}
   \Delta _{LL}\sigma  = \Delta _{LL}\hat\sigma  \times
          \Delta \hbox{pdf}_{1} \times  \Delta \hbox{pdf}_{2}.
\end{equation}
Here $\Delta _{LL}\hat\sigma $ is the corresponding helicity asymmetry in the
short-distance cross section, and $\Delta \hbox{pdf}_{i}$ is the helicity
asymmetry of the number density of partons in initial hadron $i$.
In this formula, there is an implicit sum over parton types and
an integral over the parton kinematics.
This formula is valid up to corrections that are suppressed by a
power law of the large scale.  (These are the higher twist
terms.)

A similar formula can be written when both beams are transversely
polarized.  But in many cases the numerical values of the
asymmetries are expected to be small.

When one of the beams is transversely polarized and the other is
unpolarized, QCD typically predicts \cite{KPR} that the spin asymmetry is
twist 3, i.e., $O(1/Q)$, as a consequence of helicity
conservation at the elementary vertices of the theory.  This
prediction contrasts with the notoriously large measured
asymmetries in single pion production \cite{singlepionexpt}. A
devil's advocate would argue that ``Helicity conservation is
violated whenever it has been directly measured.'' This statement
has a certain element of validity, even though there is much
quantitative evidence for the correctness of the QCD interaction
vertices, and hence for helicity conservation.


\section {Measurements of Helicity Parton Distributions}

One just has to look at other presentations at this conference to
realize the importance of measuring the helicity asymmetry of the
parton densities in the proton.  Knowing the polarization of the
sea quarks and antiquarks and of the gluon gives important
information on the spin and chiral structure of the proton.

Since neutrino scattering on a polarized target is impractical,
one can look to hadron-hadron scattering to provide much of the
information on flavor separated distributions.  One can look to
all the standard processes: Drell-Yan (both to muon pair and to W
and Z) provides a direct measurement of antiquark distributions,
direct photon, heavy flavor, and jet production probe the quark,
antiquark and gluon densities in various combinations.  Jet
production is particularly useful because of its high rate.   In
the unpolarized case, jet production is usually regarded as a
cross section predicted with the aid of other measurements.  In
the polarized case, it can be used to probe the gluon density.


\section {Transverse Polarization at Leading Twist; Fragmentation}

The distributions of transversely polarized quarks in a
transversely polarized proton are perfectly good twist-2
observables \cite{RS}, and there is an interesting range of
cross sections sensitive to them \cite{AM}.  But many spin
asymmetries that they give are likely to be small.  The
transverse spin distributions differ from the helicity
distributions because of relativistic effects in the proton wave
function, and are therefore of great interest.

It has recently been realized (or rediscovered!) that one should
investigate spin-dependent asymmetries in jet fragmentation
\cite{older,EMT,trfrag}.  These give twist-2 observables in
collisions with only one incoming beam being polarized
transversely.  The usual theorem about single-spin observables
being higher twist is overcome by using a spin sensitive
observable in the final-state---an azimuthal distribution of the
leading pion or of the leading two pions, for example.  Another
leading twist observable that has been discussed is the
polarization of a $\Lambda $ baryon at high transverse momentum.

Such measurements have the disadvantage of measuring the
transverse spin distributions only in conjunction with the spin
dependence of fragmentation.  This is also an advantage, since
the transverse-spin-dependence of fragmentation is a completely
virgin subject and depends on chiral symmetry breaking in a
situation dominated by dynamic (as opposed to static) processes.


\section {Twist-3}

There has been much theoretical work on twist-3 effects
\cite{BB,QS,JJ}.  In appropriate observables, these are
the leading twist terms. These observables include all the
simplest spin asymmetries in a collision of transversely
polarized hadron with an unpolarized hadron.

Measurements of such quantities would provide much information
about the spin structure of the proton, but are likely to be much
harder to interpret than twist 2 measurements.  Once one goes
beyond the tree graph level for the Wilson coefficients, the
number of operators involved is enormous as compared to the
twist-2 case.


\section {Outlook}

The is much interesting QCD physics to be done at a polarized
proton-proton collider.  This could be done at RHIC at a modest
incremental cost.  Both longitudinal spin and transverse spin are
of interest.

Because of this and because of the past, present and future
measurements of polarized deep inelastic scattering, there has
been a resurgence of theoretical interest.

Work needs to be done to decide on optimal detector
configurations etc.

\section*{Acknowledgments}
This work was supported in part by the U.S. Department of Energy
under grant DE-FG02-90ER-40577, by the National Science
Foundation, and by the Texas National Laboratory Research
Commission.

%


\begin {thebibliography}{AB}
%
\bibitem{RSC}G. Bunce,  J.C. Collins, S. Heppelmann, R. Jaffe,
   S.Y. Lee, Y. Makdisi, R.W. Robinett, J. Soffer, M. Tannenbaum,
   D. Underwood, A. Yokosawa, Physics World 3, 1 (1992).
\bibitem{snake}Y.S. Derbenev and A.M. Kondratenko, Sov.\ Phys.\
   Doklady 20, 562 (1976);
   Y.S. Derbenev et al., Particle Accelerators 8, 115
   (1978).
\bibitem{snaketest}A.D. Krisch, in {\em Proceedings of the Polarized
   Collider Workshop}, ed.\ J.C. Collins, S. Heppelmann and
   R. Robinett, AIP Conference Proceedings 223 (AIP, New
   York, 1991).
\bibitem{KPR}E.g., G.L. Kane, J. Pumplin, and  W. Repko,
   Phys.\ Rev.\ Lett.\ 41, 1689 (1978).
\bibitem{singlepionexpt}E704 Collaboration, D.L. Adams et al.,
   Phys.\ Lett.\ B276, 531 (1992), and references therein.
\bibitem{RS}J.P. Ralston and D.E. Soper, Nucl.\ Phys.\
   B152, 109 (1979).
\bibitem{AM}X. Artru and M. Mekhfi, Zeit.\ f.\ Phys.\
   C45, 669 (1990)
\bibitem{older}O. Nachtmann, Nucl.\ Phys.\
   B127, 314 (1977);
   R.H. Dalitz, G.R. Goldstein and R. Marshall,
   Zeit.\ f.\ Phys.\ C42, 441 (1989).
\bibitem{EMT}A.V. Efremov, L. Mankiewicz and N.A. T\"ornqvist,
   Phys.\ Lett.\ B284, 394 (1992).
\bibitem{trfrag}\sloppy J.C. Collins, S. Heppelmann, and G. Ladinsky,
   ``Measuring Transversity Densities in Singly Polarized
   Hadron-Hadron Collisions'', Penn State preprint PSU/TH/101,
   to be published in Nucl.\ Phys.\ B;
   J.C. Collins, Nucl.\ Phys.\ B396, 161 (1993).
\bibitem{BB}I. Balitsky and V. Braun,
   Nucl.\ Phys.\  B361, 93 (1991).
\bibitem{QS}J.-W. Qiu and G. Sterman,
   Phys.\ Rev.\ Lett.\ 67, 2264 (1991).
\bibitem{JJ}R.L. Jaffe and X.-D. Ji, Phys.\ Rev.\ Lett.\
   67, 552 (1991);
   X.-D. Ji, Phys.\ Lett.\ B284, 137 (1992).
\end {thebibliography}
\end {document}